\documentclass[pra, aps, twocolumn, floatfix, superscriptaddress]{revtex4}

\usepackage{graphicx, epsfig}
\usepackage{amssymb,amsmath}

\def \hf{\tfrac{1}{2}}

\def\lbc{\left[}    \def\rbc{\right]}

\DeclareMathOperator{\tr}{tr}

\newcommand{\maj}{\succ}
\begin{document}

\title{Entanglement signatures of quantum Hall phase transitions }

\author{Oleksandr Zozulya} 
\affiliation{Institute for Theoretical Physics, University of Amsterdam,
Valckenierstraat 65, 1018 XE Amsterdam, the Netherlands}

\author{Masudul Haque} 
\affiliation{Max Planck Institute for the Physics of Complex Systems
Noethnitzer Strasse 38, 01187 Dresden, Germany}

\author{Nicolas Regnault} 
\affiliation{Laboratoire Pierre Aigrain,
  Dpartement de Physique, ENS, CNRS, 24 rue Lhomond, 75005 Paris, France }

%% \author{Kareljan Schoutens} 
%% %
%% \affiliation{Institute for Theoretical Physics, University of
%% Amsterdam, 
%% %Valckenierstraat 65, 1018 XE Amsterdam,  
%% the Netherlands}

\date{\today}

%
%     Abstract
%

\begin{abstract}

  We study quantum phase transitions involving fractional quantum Hall states,
  using numerical calculations of entanglements and related quantities.  We
  tune finite-size wavefunctions on spherical geometries, by varying the
  interaction potential away from the Coulomb interaction.  
  We uncover signatures of quantum phase transitions contained in the
  scaling behavior of the entropy of entanglement between two parts of the
  sphere.
  In addition to the entanglement entropy, we show that signatures of quantum
  phase transitions also appear in other aspects of the reduced density matrix
  of one part of the sphere.

\end{abstract}

%\pacs{05.30.Pr}

%%  05.30.Pr = Fractional statistics systems, 

    %%    find keywords

\maketitle

\section{Introduction}

An exciting interdisciplinary development in recent years has been the
description of condensed matter phases using entanglement measures borrowed
from the field of quantum information theory
\cite{AmicoFazioOsterlohVedral_RMP08}.  
One very recent example is the characterization of \emph{topological order} in
fractional quantum Hall states \cite{our-prl07, our-pfaffian-entng,
LiHaldane_PRL08} using entanglement.
Topologically ordered ground states are characterized by fractionalized
excitations, degeneracies on higher-genus surfaces, and an energy gap in the
excitation spectrum above the ground state \cite{top-order_various}.  For
example, on a genus-$g$ surface, the Laughlin state at filling $\nu=1/m$ has
$m^g$ ground states.
Although a number of spin models theoretically possess topological order, the
only confirmed \emph{experimental} realizations of topological order are
fractional quantum Hall (FQH) states of two-dimensional (2D) electrons in a
magnetic field.
In this Article we focus on this most realistic class of topologically ordered
states.  FQH states have been the object of further intense attention because
of the possibility of cold-atom realizations \cite{cold-atom-FQH}, and more
recently due to quantum computation proposals based on their topological
properties \cite{topol-quantum-computing}.

In considering entanglement properties, we will focus on \emph{bipartite}
entanglement between two parts ($A$ and $B$) of the system.  This is
characterized by the reduced density matrix $\rho_A = \tr_B\rho$ of subsystem
$A$, obtained by tracing out all the $B$ degrees of freedom.  While it is
instructive to study the upper part of the eigenvalue spectrum of $\rho_A$,
the so-called \emph{entanglement spectrum}, it is also often convenient to
extract a single number from this spectrum.  For the latter purpose we will
use the entanglement entropy, $S_A = - \tr\lbc\rho_A\ln\rho_A\rbc$.

To study topological order in FQH states, Refs.\ \cite{our-prl07,
our-pfaffian-entng} exploited the novel concept of topological entanglement
entropy \cite{KitaevPreskill_PRL06, LevinWen_PRL06}, which appears in the
entanglement entropy $S_A$ between a block $A$ and the rest ($B$) of the
system.  On the other hand, Ref.\ \cite{LiHaldane_PRL08} studies the spectrum
of $\rho_A$ and extracts information relevant to the topological order from
one end of the spectrum.

A quantity characterizing a phase should also provide signatures of quantum
phase transitions leading into or away from that phase.  This idea is already
fruitful in 1D physics, where the (asymptotic) block entanglement entropy can
distinguish between gapless and gapped phases.  This allows one to pinpoint
the phase transition between gapped and gapless phase, in particular in DMRG
calculations where the block entanglement entropy is readily available.
For the case of quantum phase transitions involving topologically ordered
states, entanglement studies have been exploited for the simpler `toy' case of
Kitaev models with order-destroying additional terms
\cite{CastelnovoChamon_topol-QPT_PRB08, Hamma-etal_KitaevQPT_PRB08}.  In this
Article, we will examine the utility of entanglement calculations for studying
more `realistic' topological phase transitions, namely, transitions involving
FQH states.  FQH states being the only experimental examples of topological
order, this is an important step toward developing entanglement-based tools
that are useful for experimentally relevant situations.

We study phase transitions between FQH and non-FQH states driven by a change
in the interaction potential.  Specifically, using the Haldane pseudopotential
description of interactions projected to specific Landau levels
\cite{Haldane_PRL83}, we vary the first pseudopotential $V_1$ away from its
Coulomb potential value.  The transition we mainly focus on involves the
best-known FQH state, for fermions at filling $\nu=1/3$.  With a Coulomb
interaction between the fermions, the ground state is known to be an FQH state
topologically equivalent to the Laughlin state \cite{Laughlin_PRL83}.  If the
the first pseudopotential is reduced enough, the Laughlin state ceases to be a
good description of the ground state.  There is thus a phase transition as a
function of $V_1$ \cite{HaldaneRezayi_PRL85}.  We will present entanglement
calculations which probe this phase transition.
We also present results for filling fraction $\nu=5/2$, where the possibility
of a more intricate quantum Hall state (the Moore-Read state
\cite{MooreRead_NuclPhys91}), provides a more challenging situation.
%
%% We also present results for two other filling fractions, where the possibility
%% of more intricate quantum Hall wavefunctions, namely the Moore-Read state at
%% filling $\nu=5/2$ and the $k=3$ Read-Rezayi state at filling $\nu=13/5$,
%% provide more challenging situations.

%The second case concerns  $\nu=5/2$ filling. (or 2/5 ?)

In Section \ref{sec_background}, we summarize necessary background material on
the topological entanglement entropy $\gamma$ and on FQH wavefunctions and
transitions, and develop the scaling concepts necessary for treating phase
transitions. 
In the subsequent sections, we use three different aspects of entanglement to
probe these phase transitions.  First, we consider the entanglement entropy of
a block with the rest of the system, and track phase transitions through the
behavior of this quantity as a function of block size and system size.
%
%% The analysis of the block entanglement entropy is mostly in the context of the
%% topological entanglement entropy, which we employ for generic considerations
%% of topological phase transitions in Section \ref{sec_top-order-n-block}.
%
Since our calculations are based on finite-size wavefunctions, there are
limiting procedures involved which can be performed in different orders.
Results obtained using different limit orders are discussed in Sections
\ref{sec_original-xtrapol-method}, \ref{sec_area-law-first} and
\ref{sec_hemispheres}.

Second, we consider the top part of the reduced density matrix spectrum.  This
analysis is based on the identification of features of the entanglement
spectrum in terms of topological order and related edge physics
\cite{LiHaldane_PRL08}.  We show in section \ref{sec_entanglement-spectrum}
how the entanglement spectrum is affected when the system is driven through a
quantum phase transition.

Finally, in section \ref{sec_majorization} we use the concept of
\emph{majorization}, which is based on comparing complete reduced density
matrix spectra for two wavefunctions.  Majorization relations between reduced
density matrices obtained from condensed matter wavefunctions has been the
subject of intriguing recent studies \cite{LatorreRicoVidal_PRA05, orus_maj}.
While the full implications are not yet clear, this work adds to the growing
understanding of majorization in condensed-matter systems.

\section{Entanglement and FQH states}  \label{sec_background}

\begin{figure}
\centering \includegraphics*[width=0.85\columnwidth]{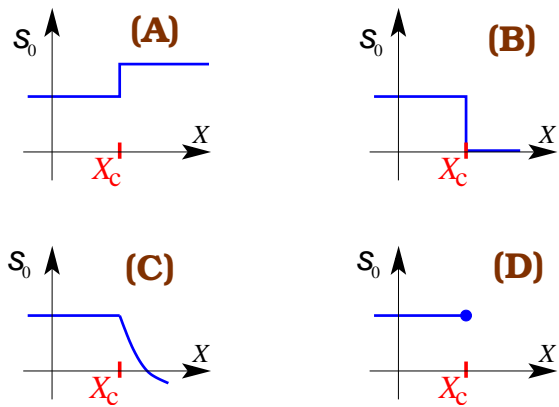}
\caption{  \label{fig_cartoon_possibilities}
(Color online)
Possible behaviors of $s_0$, defined in Eq.~\eqref{eq_s0-defn}, within a
topologically ordered phase ($X<X_{\rm c}$) and after a phase transition into
a different phase ($X>X_{\rm c}$).
}
\end{figure}

%\subsection{Topological entanglement entropy}

\subsection{Block entanglement entropy and topological phase transitions}

For 2D topologically ordered systems, an important recent result
\cite{KitaevPreskill_PRL06, LevinWen_PRL06} relates the block entanglement
entropy $S_A$ to the topological order.  The dependence of the block
entanglement entropy $S_A$ on the length $L$ of its boundary is asymptotically
linear, in accord with the ``area law''.
In addition, Refs.~\cite{KitaevPreskill_PRL06, LevinWen_PRL06} have found that
this dependence also has a topological sub-leading term:
\[
S_A(L) \xrightarrow{L\rightarrow\infty} \alpha{L} -\gamma \; .
\]
The sub-leading term is called the topological entanglement entropy
(abbreviated topological entropy) and is a constant for a given topologically
ordered phase, $\gamma=\ln\mathcal{D}$.  Here $\mathcal{D}$ is the \emph{total
quantum dimension} characterizing the topological field theory describing the
phase.

The concept of topological entropy is an important new development, because
the usual definition of topological order is often unwieldy to use in
practice for theoretical or numerical studies of such order.  The
entanglement-based characterization provides a new route for identifying
topological order and, by extension, topological phase transitions.  
After the quantity ($\gamma$) was introduced for topologically ordered
states in general, it has been explored in several specific contexts,
including quantum Hall states \cite{our-prl07, our-pfaffian-entng,
  FriedmanLevine_torus-2007, Fradkin_top-entropy-inChernSimons_JHEP08,
  MorrisFeder_arxiv0808}, quantum dimer models
\cite{FurukawaMisguich_PRB07}, and Kitaev models
\cite{CastelnovoChamon_PRB07, CastelnovoChamon_topol-QPT_PRB08,
  Hamma-etal_KitaevQPT_PRB08, CastelnovoChamon_arxiv0804,
  IblisdirPerezGarciaAguadoPachos_arxiv0806}.

We will employ considerations of both the leading linearity and the sub-leading
invariant term to characterize topological phase transitions. 
%
%\subsection{Block entanglement intercept at phase transition} \label{sec_top-order-n-block}
%
Let us consider, very generally, quantum phase transitions between a
topologically ordered phase and another phase.
For the block entanglement entropy, let us imagine that we have determined the
asymptotic relationship
\begin{equation} \label{eq_s0-defn}
S_L \quad \xrightarrow{L\rightarrow\infty} \quad  \alpha{L} \; - \; s_0
\end{equation}
where $L$ is the boundary of the block.  Note that this is not always
possible; in some 2D phases the leading term might not be purely
linear \cite{GioevKlich_PRL06, Wolf_PRL06,
  BarthelChungSchollwoeck_PRA06, RoscildeHaas_PRB06}.
Note also that the above behavior does not necessarily imply topological
order; Ref.~\cite{FurukawaMisguich_PRB07} gives an example of a
non-topological state following such a relationship with nonzero $s_0$.

If we are in a topologically ordered phase, the  negative intercept $s_0$ will
by definition be equal to the topological entropy, $\gamma = \ln\mathcal{D}$.  
Figure \ref{fig_cartoon_possibilities} shows some possibilities of what can
happen to $s_0$ as the 2D system is driven across a quantum phase transition
away from the topologically ordered state, by varying a parameter $X$ across
the critical value $X_{\rm c}$.  In the parameter range $X<X_{\rm c}$ where
the system is in the topologically ordered phase, $s_0$ is fixed at a positive
plateau ($s_0=\gamma$).

Case A shows a transition into another topologically ordered state with a
different quantum dimension; $s_0$ jumps to another constant value $\gamma'$.
The other figures show transitions to non-topological phases.  Case B shows a
transition to a gapped state which is not topologically ordered -- the
intercept $s_0$ drops to zero. (Ref.~\cite{CastelnovoChamon_topol-QPT_PRB08}
treats an example.)  Case C shows a transition into a non-topological phase,
in which $s_0$ is nonzero but not constant.  Finally, Case D shows a
transition into a state where the leading term in the asymptotic behavior of
$S_A(L)$ is not linear, so that $s_0$ is undefined.

\subsection{FQH wavefunctions on spheres}

For finite-size studies of fractional quantum Hall physics, spherical and
toroidal geometries are particularly popular when one wants to avoid
complications due to the edge.  We set notation by providing a rapid review of
FQH wavefunctions on spheres \cite{Haldane_PRL83, HaldaneRezayi_PRL85}.  We
study $N$ electrons on a sphere of radius $R$, subjected to the magnetic field
$B$ provided by a magnetic monopole at the center of the sphere.
The Dirac quantization condition requires the number of fluxes $N_{\phi} = 4
\pi R^2 B/ \phi_0$ to be integral, where $\phi_0 = h/e$ is the quantum of
flux.  If we measure length in units of the magnetic length $l_B =
\sqrt{\hbar/e B}$, then the quantization condition is $R = \sqrt{N_{\phi}/2}$.
The $N_{\phi}+1$ Landau-level orbitals are labeled either $l=0$ to $N_{\phi}$
or $L_z=-L$ to $+L$.  The wavefunctions in spherical coordinates are given by
\[
| \Psi_l(\theta, \varphi)|^2 = \left(\cos \frac{\theta}{2}
  \right)^{2N_{\phi}-2l} \left(\sin \frac{\theta}{2}\right)^{2l}  \; .
\] 
There is no dependence on the azimuthal angle; the orbitals are each localized
around a ``circle of latitude'', with the $l=0$ orbital localized near the
``north pole.''  
Basis states for the $N$-electron wavefunctions are expressed in terms of the
occupancies of these orbitals.  

%in the 'Landau gauge' $A_{\theta} = 0,~ A_{\varphi} = B R \cot \theta$.
%

To define entanglements between two parts of the sphere, one has to first
partition the sphere into $A$ and $B$.  We define block $A$ to be the first
$l_A$ orbitals, extending spatially from the north pole out to some latitude,
\emph{i.e.}, including orbitals $l=0$ through $l=l_A-1$.

While neighboring orbitals do overlap, it is natural to think of the location
where their amplitudes are equal, $|\Psi_l(\theta)|^2 =
|\Psi_{l+1}(\theta)|^2$, as the ``boundary'' between the orbital $l$ and the
orbital $(l+1)$.  This happens at angle
\[
 \tan^2 \frac{\theta_l}{2} = \frac{l+1}{N_{\phi}-l} \; .
\]
The boundary between partitions $A$ and $B$ is thus a circle of latitude at
polar angle $\theta_{l_A-1}$.  The boundary length is
\[
 C_{l_A}(N) ~=~ 2 \pi R \sin \theta_{l_A-1} 
 ~=~  \frac{4 \pi R \sqrt{ l_A (N_{\phi}+1 - l_A)}}{N_{\phi}+1} \; .
\]

%\subsection{The shift}

In finite-size spherical geometries, FQH states do not appear exactly at the
filling fraction $\nu$.  Instead, the number of orbitals $(N_{\phi}+1)$ is
related to the number of particles $N$ by:
\[
N_{\phi} = N \nu^{-1} -S ~,
\]
where the shift $S$ is an integer determined by the FQH state
\cite{Wen_shift}.  Although $S$ is insignificant in the thermodynamic limit,
it can create an ``aliasing'' problem in numerical studies: different FQH
states can compete at the same finite values of $N$ and $N_{\phi}$.

\subsection{Exact diagonalization}

Our analysis is based on numerical exact diagonalization of Coulomb-like
Hamiltonians projected onto appropriate Landau levels (lowest Landau level for
$\nu=1/3$ and the second Landau level for $\nu=5/2$).
% and the next one for $\nu=5/2$ and $\nu=13/5$).  
%
We employ the spherical geometry described above.
For this work, we have calculated $\nu=1/3$ wavefunctions up to 14 fermions
and $\nu=5/2$ wavefunctions up to 20 fermions.  To give an idea of the scale
of these calculations, we note that for the $N=20$ wavefunctions at $\nu=5/2$
the Hilbert space dimension is 193498854, while for the $N=14$ wavefunctions
at $\nu=1/3$, it is 129609224.  Dimensions are given here after reduction
using the discrete symmetry under global $L_z$ flip.

We used the Lanczos algorithm to compute the system ground state.
Calculations were performed on a PC cluster with 24 cores (AMD Opteron 265)
and 48 Gbytes of memory.  A Lanczos iteration typically requires up to 17h of
cpu time for the largest Hilbert spaces.
In our study of phase transitions, calculations are particularly
time-intensive when the system is in or near a gapless phase, in which case
the Lanczos procedure requires a larger number of iterations to converge.

\begin{figure}
 \includegraphics*[width=0.99\columnwidth]{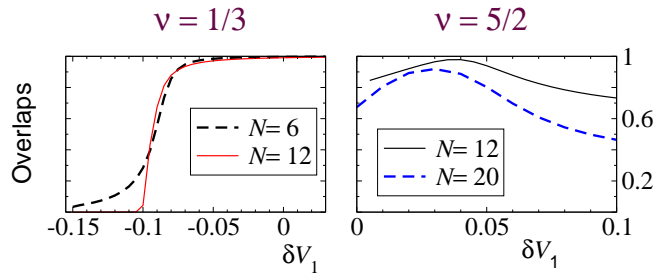}
\caption{  \label{fig_overlaps}
(Color online)
Overlaps of ground state wavefunctions at interactions relevant to fillings
1/3 and 5/2, respectively with Laughlin and Moore-Read wavefunctions.
%
%% Overlaps of ground state wavefunctions at interactions relevant to fillings
%% 1/3, 5/2, and 13/5, respectively with Laughlin ($\nu=1/3$), Moore-Read
%% ($\nu=1/2$) and Read-Rezayi ($\nu=13/5$) wavefunctions.  
%
A large-$N$ and a small-$N$ case are shown for each fraction.
}
\end{figure}

\subsection{Phase transitions}

For FQH systems, the natural interaction parameter to vary is one of the
Haldane pseudopotentials $V_m$ obtained by decomposing the interaction
potential into channels specified by the relative angular momentum $m$ of the
interacting particles \cite{Haldane_PRL83, HaldaneRezayi_PRL85}.
We use interaction potentials whose $m=1$ pseudopotential $V_1$ is changed,
while the other $V_m$ are fixed at the Coulomb value for that Landau level.
We calculate and present wavefunction properties as a function of
${\delta}V_1\propto(V_1-V_1^{\rm coul})$.
Tuning of $V_1$ loosely represents variable aspects of FQH experiments, such
as the thickness of the quantum well where the 2D electron gas resides.

For the $\nu=1/3$ case, the exact Laughlin state is obtained for
${\delta}V_1\rightarrow{+\infty}$ (short-range interaction), and continues to be a
good description of the state for the Coulomb potential, ${\delta}V_1=0$.  However,
for negative ${\delta}V_1\propto(V_1-V_1^{\rm coul})$, there is a quantum phase
transition at some ${\delta}V_1 = {\delta}V_{1c} < 0$ into a non-FQH incompressible state
\cite{HaldaneRezayi_PRL85}.  The overlap plots of Fig.\ \ref{fig_overlaps}
(left panel) shows the transition to be somewhere between ${\delta}V_1 = -0.065$ and
${\delta}V_1 = -0.11$.  The non-FQH state for ${\delta}V_1 < {\delta}V_{1c}$ presumably has
charge-density-wave or Wigner-crystalline order \cite{HaldaneRezayi_PRL85};
the details are not important for our purposes.

For $\nu=5/2$, the overlaps (figure \ref{fig_overlaps}) suggest that the
Moore-Read state is stable in some window of slightly positive ${\delta}V_1$ (around
${\delta}V_1\sim0.03$), and that there are transitions on either side of this phase
to non-FQH phases \cite{Morf_PRL98, RezayiHaldane_PRL00}.
%
%% Figure \ref{fig_overlaps} (right panel) indicates a similar picture for the
%% so-called $k=3$ Read-Rezayi state at $\nu=13/5$ \cite{ReadRezayi_PRB99}.
%
The non-FQH phases are possibly a striped charge-density-wave phase on the
left of the Moore-Read region and a composite fermi sea on the right
\cite{RezayiHaldane_PRL00}.

\subsection{Extrapolation for $s_0$} \label{sec_extrapolation-comments}

The definition of the topological entanglement entropy $\gamma$, or the
quantity $s_0$ in equation \ref{eq_s0-defn}, involves two extrapolations: (1)
to the thermodynamic limit, $N\rightarrow\infty$, and (2) to the asymptotic
limit of the size of the $A$ block, i.e. $l_A\gg{1}$.  This \emph{double
scaling limit} can be approached via different possible extrapolation paths in
the $(N,l_A)$ space.
To describe the different extrapolation methods, we first rewrite the relation
\eqref{eq_s0-defn} for finite $N$ and $l_A$:
\begin{equation}  \label{eq_S-vs-circum}
 S_{l_A}(N) ~=~ s_1 C_{l_A}(N) - s_0(N)
\end{equation}
We have ignored $l_A$-dependence of $s_0$ and both $N$- and $l_A$-dependence
of $s_1$.  Our experience with numerical wavefunctions indicate that these
dependences are generally weak, at least for model FQH wavefunctions.
Note that, even for topologically ordered model FQH states, the block
entanglements $S_{l_A}$ are not linear in $\sqrt{l_A}$ for finite $N$
(\emph{c.f.}\ Fig.~1 in Ref.~\cite{our-prl07}).  The reason is that
$\sqrt{l_A}$ is proportional to the square root of the area of the $A$ region,
which is not proportional to the circumference $C_{l_A}(N)$ in the curved
geometry of the sphere.

In Sec.~\ref{sec_original-xtrapol-method}, we consider the method used in
Refs.~\cite{our-prl07, our-pfaffian-entng}, namely, performing the
$N\rightarrow\infty$ extrapolation first for each $l_A$, and then using the
resulting $S_{l_A}(\infty)$ versus $C_{l_A}(\infty)\sim\sqrt{l_A}$ to extract
$s_0(\infty)$ or $\gamma$.

In Sec.~\ref{sec_area-law-first} we consider the extrapolation procedure in
reverse order, namely, first extracting $s_0(N)$ from finite-size $S_{l_A}$ versus
$C_{l_A}$ dependences, and then taking the $N\rightarrow\infty$ limit.

In Sec.~\ref{sec_hemispheres}, we take the block $A$ to be half the system
(which is a natural choice in a spherical geometry), $l_A = l_A^* =
\hf(N_{\phi}+1)$.  Examining the $S_{l_A^*}(N)$ versus $C_{l_A^*}(N)$ may be
regarded as taking the two limits simultaneously.

\subsection{The entanglement spectrum}

Clearly, the complete spectrum of the reduced density matrix $\rho_A$ of a
subsystem $A$ contains more information than any one number (such as the
entropy of entanglement $S_A$) extracted from this spectrum.  Extraction of
additional information from the complete spectrum has been reported in several
other condensed-matter contexts.

Ref.~\cite{LiHaldane_PRL08} has empirically shown that, for FQH wavefunctions
on a sphere, the spectrum of the reduced density matrix of one hemisphere can
be related to the conformal field theory (CFT) describing the 1D edge of the
FQH state.  We exploit this notion in Sec.\ \ref{sec_entanglement-spectrum} to
show how quantum phase transitions appear in the so-called entanglement
spectrum.

Another way of exploiting the complete spectrum of $\rho_A$ is via
\emph{majorization} comparisons, which we use in Sec.\ \ref{sec_majorization}.

\begin{figure}
 \includegraphics*[width=0.95\columnwidth]{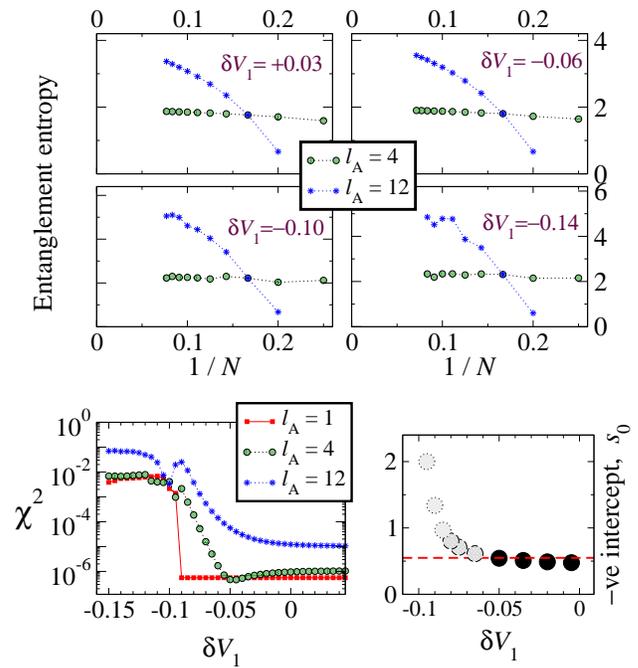}
\caption{  \label{fig_fixed-lA-scalings}
(Color online)
Top four: $S_{l_A}(N)$ versus $1/N$ for several ${\delta}V_1$ values, displaying how
the smooth behavior of the these curves disappear.  
Bottom left: quality of fit ($\chi^2$) to a simple function, $c_0+c_1/N^2$.
Bottom right: Attempt to extract $s_0$.  The quantity is less and less
meaningful for more negative ${\delta}V_1$ (discussion in text).  
The topological value $\hf\ln3 \approx 0.55$ is shown by dashed horizontal line.
}
\end{figure}

\section{Extrapolation for each block size} \label{sec_original-xtrapol-method}

We attempt to employ the extrapolation method of Refs.~\cite{our-prl07,
our-pfaffian-entng}, first extrapolating $N\rightarrow\infty$ for fixed values
of $l_A$.
From the $S_{l_A}(N)$ versus $1/N$ plots in figure
\ref{fig_fixed-lA-scalings} (top 4 panels), we note that the
extrapolation starts to lose meaning as one reduces $V_1$ beyond the
presumed transition.  This is made quantitative by estimating
goodness-of-fit of the $S_{l_A}(N)$ versus $1/N$ data to simple
functions.  The plotted $\chi^2$ estimates are obtained from trying
$S_{l_A}(N) = c_0+c_1/N^2$; any other reasonable function gives
similar results.  This $\chi^2$ versus ${\delta}V_1$ curve can be
regarded as one entanglement signature of the phase transition.

The jump in $\chi^2$ gets sharper for smaller $l_A$.  In particular, for
$l_A=1$ it is distinctly localized at ${\delta}V_1\sim{-0.1}$.  This appears to
provide a sharp estimate for the location of the transition.  
One tempting interpretation is that, since the system is large
compared to a single orbital, $S_{l_A=1}(N)$ for moderate $N$ is
already a good indicator of the thermodynamic limit; hence the sharp
jump.  
% However, further understanding is needed to evaluate this idea.

One could still estimate $S_{l_A}(N\rightarrow\infty)$ extrapolates,
disregarding the ``noise'' in the $S_{l_A}(N)$ versus $1/N$ datapoints.  The
resulting $S_{l_A}(\infty)$ points, plotted against $\sqrt{l_A}$, gives a line
whose intercept is by definition $-s_0$.  The estimates of $s_0$, thus
obtained, move away from the expected value of $\gamma =
\ln\sqrt{3}\sim{0.55}$ around the transition point.
However, it is important to note that in the region
${\delta}V_1\lesssim{-0.06}$, the estimates have relatively little meaning.
First, there is the increasing scatter in the $S_{l_A}(N)$ versus
$1/N$ data, which makes the $S_{l_A}(\infty)$ values unreliable.
Second, we find that the $S_{l_A}(\infty)$ versus $\sqrt{l_A}$ curves
are more and more curved as one increases $-{\delta}V_1$ further away from
the Laughlin state.
Given the uncertainties, we do not attempt to estimate error bars for $s_0$.
The point here is to note how $s_0$ loses meaning, in the large-$(-{\delta}V_1)$
region for which we have used shaded symbols in figure
\ref{fig_fixed-lA-scalings}.
Referring back to figure \ref{fig_cartoon_possibilities}, the situation we
have should be regarded more like case D and not like case C.
%

%% Note that neither the overlap not the $s_0$ calculations in
%% Fig.~\ref{fig_fixed-lA-scalings} pinpoint the transition point ${\delta}V_{1c}$
%% very sharply.

\section{Dependence on circumference and area law}  \label{sec_area-law-first}

\begin{figure}
\centering
 \includegraphics*[width=0.95\columnwidth]{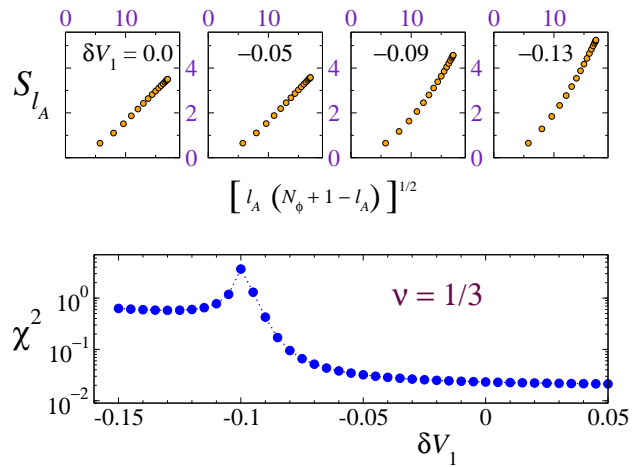}
\caption{  \label{fig_EntVsCircum}
(Color online)
Top panels: entanglement entropy plotted against circumference of $A$ block; $\nu=1/3$, $N=12$.
The linearity is progressively destroyed as one reduces the pseudopotential
$V_1$.  Lower panel: $\chi^2$ for linear fits to the $S_{l_A}$ versus
circumference data.
}
\end{figure}

We now consider extrapolations in reverse order, i.e., first find a negative
intercept $s_0(N)$ for each $N$ by looking at the large-$l_A$ behavior at that
$N$, and only afterwards consider the $N$-dependence.
For topologically ordered model FQH states, the block entanglements
$S_{l_A}(N)$ have good linear dependences on the circumference $C_{l_A}(N)
\propto \sqrt{l_A(N_{\phi}+1-l_A)}$, for any fixed $N$.  

Here we are of course interested in realistic wavefunctions, \emph{i.e.}, the
ground states of Coulomb-like potentials.
In figure \ref{fig_EntVsCircum}, $S_{l_A}$ versus $C_{l_A}$ plots are
shown for several ${\delta}V_1$, for $\nu=1/3$ states of 12 particles.
It is interesting to note that the plots remain smooth as
$-{\delta}V_1$ is increased past the transition; however they acquire
curvature.
As the dependence deviates from the linearity of model wavefunctions,
extracting $s_0$ loses meaning.  We do not attempt estimating $s_0$.  Instead,
the deviation from linearity is shown through $\chi^2$ measures.
% , for two values of $N$.
The rise in $\chi^2$ again represents the quantum phase transition.

Note that this deviation from linearity does not necessarily constitute a
violation of the area law, which is a statement about the thermodynamic limit.
The deviation could instead indicate a stronger $l_A$-dependence of the
quantity $s_1$ in Eq.~\eqref{eq_S-vs-circum}.

\section{Entanglement between equal hemispheres} \label{sec_hemispheres}

\begin{figure}
\centering
 \includegraphics*[width=0.98\columnwidth]{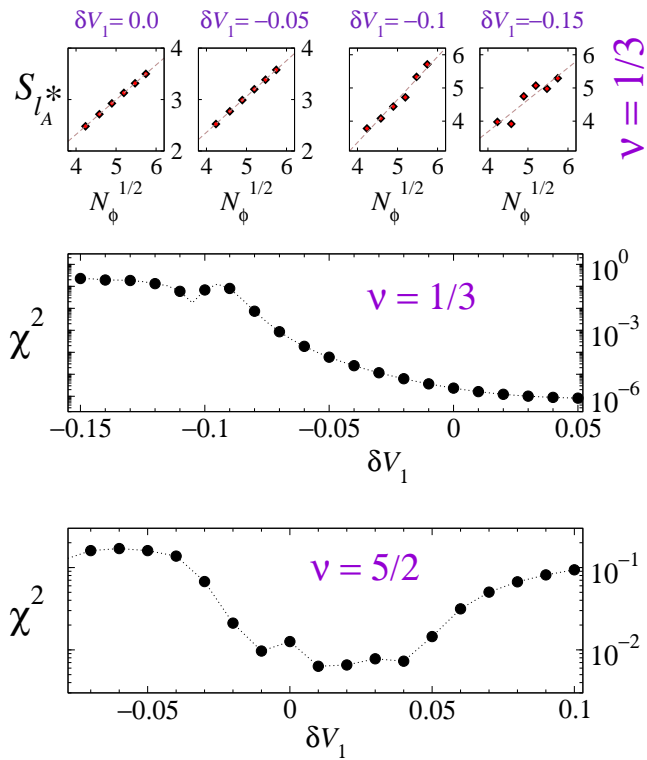}
\caption{  \label{fig_hemisphere-chisq}
(Color online)
Top row: entanglement entropies for hemisphere partitioning, plotted against
sphere radius;  $\nu = 1/3$.  Center row shows quality of linear fit as function
of pseudopotential ${\delta}V_1$ for $\nu = 1/3$.  Bottom row shows $\chi^2$ for $\nu
= 5/2$.
}
\end{figure}

We now consider entanglement between equal hemisphere-shaped $A$ and $B$
blocks, \emph{i.e.}, we use $l_A = l_A^* = (N_{\phi}+1)/2$.  According to
Eq.~\eqref{eq_S-vs-circum}, $S_{l_A^*}(N)$ should depend linearly on the
circumference $C_{l_A^*}(N) \equiv 2 \pi R \propto \sqrt{N_{\phi}}$.  For the
exact Laughlin state, an example is shown in Fig.~1 (inset) of
Ref.~\cite{our-prl07}.  Here, we explore the fate of this linearity as a
function of ${\delta}V_1$.

The top row in figure \ref{fig_hemisphere-chisq} plots $S_{l_A^*}$ against
$\sqrt{N_{\phi}}$ for several $V_1$ values at $\nu=1/3$.
The points acquire scatter as $-{\delta}V_1$ is increased, until it becomes
meaningless to extract the intercept; the situation is similar to Section
\ref{sec_original-xtrapol-method} (figure \ref{fig_fixed-lA-scalings}).  This
process is indicated by the $\chi^2$ curve (center panel) signifying the
goodness of the linear fit.

The $\nu=5/2$ case (figure \ref{fig_hemisphere-chisq} bottom panel) similarly
shows that the scatter is low (so that a linear $S_{l_A^*}$ versus
$\sqrt{N_{\phi}}$ fit is meaningful) only in the window of ${\delta}V_1$ where the
Moore-Read state describes the physics of the ground state.

A technical note: for even $N_{\phi}$, the number of orbitals is odd and it is
impossible to divide the sphere precisely into halves.  In this case we keep
$l_A^* = N_{\phi}/2$ orbitals in block $A$, which nearly divides the sphere
into halves, and use values $S_{l_A^* = N_{\phi}/2}$ for the entropy and
$C_{l_A^* = N_{\phi}/2}(N)$ for the circumference.

\section{Entanglement spectrum and entanglement gap} \label{sec_entanglement-spectrum}

\begin{figure}
\centering
 \includegraphics*[width=0.98\columnwidth]{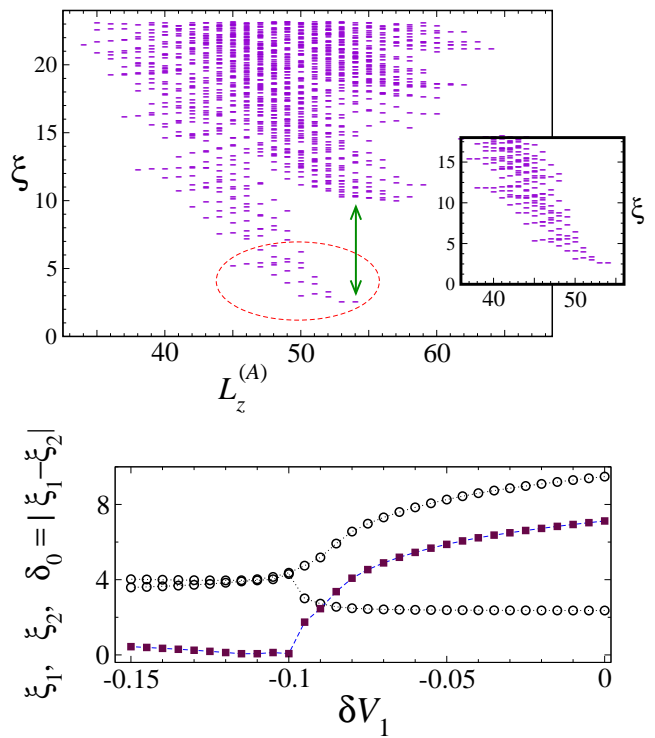}
\caption{  \label{fig_entanglement-gap}
(Color online)
Top panels: entanglement spectrum, $\nu=1/3$, $N=12$, block $A$ containing
$l_A=17$ orbitals and $N_A = 6$ fermions.  Main plot: ground state for
${\delta}V_1=0.04$.  Ellipse indicates the most prominent ``conformal'' part of the
spectrum.  
Arrow indicates the ``entanglement gap'' $\delta_0$ between CFT and non-CFT
parts of the spectrum.  
Inset shows exact Laughlin state, which has no higher-lying non-CFT
part.
Lower panel: Empty dots show two lowest levels at $L_z^{(A)}=54$, 
% where the ``vacuum state'' of the CFT appears,
plotted against ${\delta}V_1$.
Filled squares show ``entanglement gap'', the difference of the two lowest
levels.
}
\end{figure}

Following Ref.\ \cite{LiHaldane_PRL08}, we introduce the "entanglement
spectrum" $\xi$ as $\lambda_i = \exp(-\xi_i)$, where $\lambda_i$ are
eigenvalues of the reduced density matrix $\rho_A$ of one hemisphere.
The eigenvalues can be classified by the number of fermions $N_A$ in
the $A$ block, and also by the total ``angular momentum'' $L_z^{(A)}$
of the $A$ block.
It was argued \cite{LiHaldane_PRL08} that the low-lying spectrum $\xi_i$ of
the reduced density matrix for fixed $N_A$, plotted as a function of $L_z^{(A)}$,
should display a structure reflecting the conformal field theory (CFT)
describing the edge physics.  In figure \ref{fig_entanglement-gap} this ``CFT
spectrum'' is marked with an ellipse.
For interactions at which the FQH state provides a good description of
the physics, the CFT spectrum is well-separated by a gap from a higher
``non-CFT'' part of the spectrum.

As in Ref.\ \cite{LiHaldane_PRL08}, we denote the gap between the
lowest two $\xi_i$, at the $L_z^{(A)}$ value where the
highest-$L_z^{(A)}$ member of the CFT spectrum occurs, as $\delta_0$.
In figure \ref{fig_entanglement-gap}, this is the gap between the
lowest two states at $L_z^{(A)}=54$ (marked by arrow).
We study what happens to the spectrum as we tune the interaction away
from the FQH state across a quantum phase transition.  We quantify the
change of the spectrum in terms of the quantity $\delta_0$, defined
above.  (The quantities $\delta_{1,2}$ defined in Ref.\
\cite{LiHaldane_PRL08}, the gaps at other $L_z^{(A)}$ values, are
expected to have similar dependence on ${\delta}V_1$.)

In figure \ref{fig_entanglement-gap} (lower panel), we plot $\delta_0$
as a function of the pseudopotential ${\delta}V_1$ for the $\nu=1/3$
case.  This clearly shows a dramatic decrease of the "entanglement
gap" around the region of the phase transition.  The two levels in
question are also individually plotted with open dots; there is a
level crossing around ${\delta}V_1\sim{-0.1}$.
We note that for values of ${\delta}V_1<-0.1$ the CFT-like structure of the
entanglement spectrum is lost so it is no longer meaningful to think of
$\delta_0$ as the gap between CFT and non-CFT energy levels.  A similar
picture is observed for Moore-Read wavefunctions \cite{Haldane_unpublished}.

\section{Majorization} \label{sec_majorization}

\begin{figure}
\centering 
\includegraphics*[width=0.98\columnwidth]{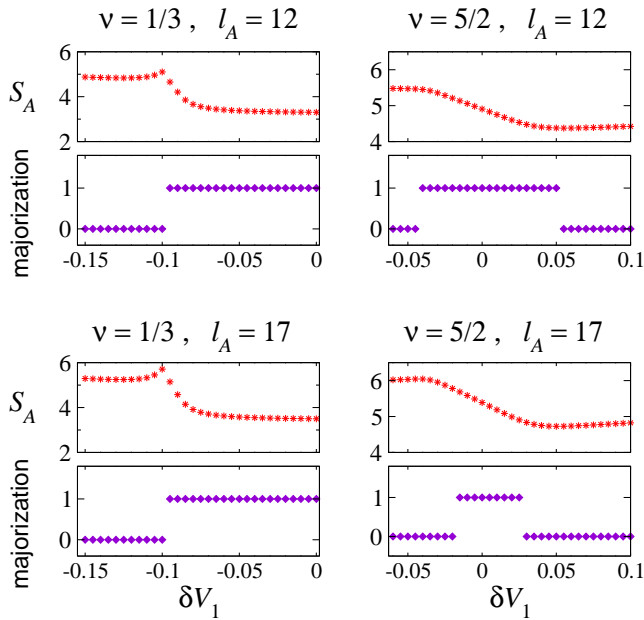}
\caption{  \label{fig_majorization}
(Color online.)
Left panels: Entanglement entropy and majorization plotted against ${\delta}V_1$ for
$N=12$, $\nu=1/3$.  Right panels: same for $N=18$, $\nu=5/2$.
Majorization at some ${\delta}V_1$ is taken to be $1$ if the corresponding $\rho_A$
spectrum majorizes a spectrum at an adjacent value of ${\delta}V_1$; otherwise it is
set to $0$.
%
%% Majorization is considered to be violated if $\min(T_k)< - \epsilon$, where
%% $\epsilon$ is the maximum numerical error in eigenvalues.
%
}
\end{figure}

The concept of majorization involves comparison of two complete
spectra.  In the context of condensed-matter applications, it
generally involves the comparison of two reduced density matrix
spectra corresponding to the ground states of two Hamiltonians with
slightly different parameters \cite{LatorreRicoVidal_PRA05, orus_maj,
  calabrese_maj}.

To define majorizaton, we consider two sets of $n$ real elements
$\{\lambda_i\}$ and $\{\mu_i\}$ sorted in decreasing order and  satisfying $\sum_i
\lambda_i = \sum_i \mu_i =1$. 
One says that the set $\{\lambda_i\}$ majorizes the set $\{\mu_i\}$ if
\[
\forall k \in \{1, \dots, n\}: \qquad \sum_{i=1}^k \lambda_i \geq \sum_{i=1}^k
\mu_i \; .
\]
This relationship is often expressed as $\lambda \maj \mu$.
%
%If neither $\lambda \maj \mu$ nor $\mu \maj \lambda$ then I call 
%sets $\lambda$ and $\mu$ as \textbf{incomparable}.
%
Obviously, if $\lambda \maj \mu$ then $S(\lambda) \leq S(\mu)$, where $S(x) =
- \sum_i x_i \ln x_i$ is the von Neumann entropy. 

In 1D quantum systems, the spectrum of the reduced density matrix
for a spatial block has been argued \cite{LatorreRicoVidal_PRA05, orus_maj,
calabrese_maj} to become more majorized as one moves along a renormalization
group (RG) flow, away from an RG fixed point.
As of now, there are no established general results for 2D
quantum states. 

In figure \ref{fig_majorization}, we examine majorization of reduced density
matrices between ground states at different values of the pseudopotential
$V_1$.  We show results for the $\nu=1/3$ system with $N=12$ particles, and
the $\nu = 5/2$ system with $N=18$ particles.  
For the $\nu=1/3$ system, we find that the eigenspectrum is continuously
majorized as ${\delta}V_1$ decreases down to the value ${\delta}V_1=-0.10$,
i.e., down to the phase transition region.  In this region, the eigenspectrum
of $\rho_A$ for each ${\delta}V_1$ majorizes the $\rho_A$ eigenspectrum at a
smaller (more negative) ${\delta}V_1$. 
For ${\delta}V_1 \leq -0.1$, the
$\rho_A$ spectra are not majorized.  This result is robust for different sizes
($l_A$) of the $A$ block.
A similar picture emerges for the $\nu = 5/2$ system; majorization occurs in a
region roughly corresponding to where the ground state has the structure of
the Moore-Read state.  However, the effect is more fragile, \emph{e.g.}, the
extent of ${\delta}V_1$ values where the majorization is found, depends on the
partition size ($l_A$) used. 
%
%% It is worth to note that one
%% has to be careful in verifying the majorization due to numerical roundoff
%% errors introduced during the computation. If a typical numerical error is
%% $\epsilon$ then we say that $\lambda \maj \mu$ if $\forall k \quad T_k \equiv
%% \sum_{i=1}^{k} (\lambda_i - \mu_i) >-\epsilon$.
% 
Note for $\nu=1/3$ filling that $S_{l_A}$ has a kink near the transition
point.  No such feature is seen for the $\nu=5/2$ case, neither in
$S_{l_A}(V_1)$ not in its derivative.

To summarize, we have demonstrated for $\nu=1/3$ that the region where
majorization occurs coincides dramatically with the region where the system is
in an FQH state; we no longer find majorization away from this phase.  The
situation is similar but less clear for $\nu=5/2$.  A full understanding is
lacking at the moment, but several intriguing speculations present themselves.
Most obviously, it is tempting to think of majorization being an indicator of
quantum phase transitions.  Second, since the reduced density matrix of a
block of the sphere contains information about the physics of the quantum Hall
edge \cite{LiHaldane_PRL08}, it is possible that our majorization results can
be interpreted in terms of the evolution of the edge as a function of
${\delta}V_1$.

\section{Conclusions}

We have presented a numerical study of how quantum phase transitions involving
fractional quantum Hall states are manifested in entanglement measures and
related quantities. 
We used three extrapolation methods to examine the double-scaling limit where
the block entanglement intercept ($s_0$ in equation \ref{eq_s0-defn}) is
defined.  We showed that the breakdown of these extrapolation procedures
signals the quantum phase transitions away from the topologically ordered
incompressible FQH states.  In addition, the entanglement spectrum was used in
more detail in two different ways to characterize the quantum phase
transitions, exploiting recently-developed concepts of CFT spectrum and
majorization.

Our work opens up several new open questions and directions.  

Since we have explored topological phase transitions on spherical
geometries only, it would be instructive to look for analogous
signatures on a toroidal geometry.  Entanglement scaling behaviors are
known in much less detail for FQH states on the torus, and further
investigations are clearly needed.

Second, our results on majorization invite a thorough investigation of
this concept, in general for two-dimensional systems and in particular
both for 2D phase transitions and for topologically ordered states.
It would appear that the knowledge necessary for putting our findings
in context does not yet exist.  The study of majorization, and the
upper part of the reduced density matrix spectrum, also raises the
question of other signatures of topological phase transitions one
might yet extract from the full $\rho_A$ spectrum.

Finally, for topologically ordered states, it is promising to explore our
three extrapolation methods, for extracting the topological quantity $\gamma$.
In previous work with FQH states on spherical geometries \cite{our-prl07,
our-pfaffian-entng}, the focus has been on the first extrapolation procedure
(section \ref{sec_original-xtrapol-method}).  If the topological entanglement
entropy is to be used to identify intricate FQH states and their CFT's,
improved methods of calculating $\gamma$ are vital.  Alternate extrapolation
methods is one direction that needs to be explored in this regard.

\acknowledgments

We acknowledge many discussions with and collaborative work with Kareljan
Schoutens, and helpful discussions with B.~A.~Bernevig, P.~Calabrese,
C.~Chamon, F.~D.~M.~Haldane, E.~H.~Rezayi.
OZ and MH thank the European Science Foundation ({\sc instans} programme) for
travel grants enabling parts of this work.

\end{document}